\documentclass[vecphys]{svmult}

\usepackage{makeidx}         
\usepackage{graphicx}        
\usepackage{multicol}        
\usepackage[bottom]{footmisc}

\begin{document}

\title*{A Novel Precise Method for Correcting the Temperature in Stellar Atmosphere
 Models}
\titlerunning{Temperature Correction in Stellar Atmospheres}
\author{Octavio Cardona\inst{1}\and   Lucio Crivellari\inst{2,3}\and   Eduardo Simonneau\inst{4}}
\authorrunning{O. Cardona et al.}
\institute{
Instituto Nacional de Astrof\'isica \'Optica y Electr\'onica. 
Luis Enrique Erro No.1. Tonantzintla, Puebla, M\'exico
\texttt{ocardona@inaoep.mx}\and INAF-Osservatorio Astronomico di Trieste, 
Trieste, Italy\and
Instituto de Astrof\'\i sica de Canarias, La Laguna (Tenerife), Spain
\texttt{luc@iac.es}
\and CNRS, Institut d'Astrophysique, Paris, France }
%
%
\maketitle

\begin{abstract}

	A mayor problem that arises in the computation of stellar atmosphere models is the self consistent determination of the temperature distribution via the constraint of energy conservation. The energy balance includes the gains due to the absorption of radiation:
$\int \chi(\nu) J(\nu) d\nu$, and the losses due to emission: $\int \chi(\nu) S(\nu) d\nu\ .$ It is well known that within each one of the two above integrals the part 
corresponding to spectral ranges whose opacity $\chi(\nu)$ is huge can overcome by 
many orders of magnitude the part that corresponds to the remaining frequencies. On
the other hand, at those frequencies where $\chi(\nu)$ is very large, the mean 
intensity $J(\nu)$ of the radiation field shall be equal, up to many significant 
digits, to the source function $S(\nu)$ and consequently to the Planck function 
$B(\nu,T)$. Then their net share to the energy balance shall be null, albeit 
separately their contributions to the gain and loss integrals are the most  
important numerically. Thus the spectral range whose physical contribution to the 
overall balance is null will dominate numerically both sides of the energy balance 
equation, and consequently the errors on the determination of $J(\nu)$ and
$S(\nu)$ at these frequencies will falsify the balance.
	It is possible to circumvent the numerical problem brought about by the 
foregoing circumstances by solving the radiative transfer equation for the variable
$I({\bf{n}},\nu) - S(\nu)$, instead of the customary intensity $I({\bf{n}},\nu)$. 

    We present here a novel iterative algorithm, based on iteration  factors already employed by us with 
success, which makes it possible a fast correction of the temperature by computing directly the difference 
between the radiative losses and gains at each step of the iterations.
\end{abstract}

\section{A Two-Block Iterative Scheme for Modeling Stellar Atmospheres}
\label{sec:1}

The distinctive feature of our algorithm for an iterative temperature correction
procedure comes from the organization into two fundamental macro-blocks of the ensemble
of the mathematical equations which represent the physical phenomena that occur in a 
stellar atmosphere.  Within each block a 
self-consistent physical problem is solved by assuming that the current values of some
input variables are known, so that the updated values of the output variables of the 
block can be obtained. 
	
The first macro-block (the {\it{structural}} one) consists of the equations 
that express the conservation of mass
and momentum. The latter, together with the equation of state, are the backbone of
 the {\it{''hard ''}} structure of the model. If we assume that the temperature 
distribution
$T(r)$ is given, 
we can deal with the equations of this block in a quite easy way, and get the 
distribution of the pressure $P(r)$ and density $\rho(r)$, hence the values of all the 
relevant thermodynamical and transport coefficients. All these values depend of course
on the initial temperature distribution assumed.

Once we know the hard structure, namely $P(r)$ and $\rho(r)$, as well as the ensemble
of the thermodynamical and transport coefficients at each point of the physical 
system, we can tackle the radiative transfer (RT) equations, coupled
through the constraint of the conservation of energy, in order to obtain with
relative ease a new temperature distribution. These equations, together with the energy
conservation constraint, form the second macro-block (the {\it{energetic}} one). 
The two
macro-blocks are coupled both physically and mathematically {\it{via}} the equation of 
state. However this coupling is not very strong, so each block can be treated
mathematically in an independent way by means of a sequential procedure that is 
iterated till the required numerical convergence.
According to our experience, such an iterative 
procedure quickly converges (in about ten iterations), and numerical 
difficulties never show up. An exhaustive presentation of the algorithm can be found in 
\cite{Simonneau88}.

\section{The Iterative Scheme for the Temperature Correction}
\label{sec:2}

The aim of this work is to discuss the numerical treatment of the second macro-block, 
given the hard structure of the system. Such a procedure is usually called the 
{\it{temperature correction}}, because it is matter of determining - also iteratively -
the temperature distribution that satisfies simultaneously both the RT equations and 
the relevant 
constraint of total energy conservation, through the correction of a previous 
temperature distribution $T^{\ i}(r)$, assumed to be close to the required one. 
For the sake of an easier presentation, we 
will consider here the instance, in which energy is transported only by radiation. 
It is simple to generalize the algorithm in case that both radiative and convective 
transport are important, as shown in~\cite{Crivellari91}. Likewise, we will
consider the paradigm case of a model atmosphere under the simplifying hypotheses of
hydrostatic and local thermodynamical equilibrium.

The RT process is described 
by means of a set of kinetic equations, one for each of the specific intensities  
$I({\bf{n}},\nu)$ employed for the representation of the radiation field. The 
corresponding sink terms, {\it{i.e.}} the specific radiative energy removed at each 
point, are assumed to be proportional to $I({\bf{n}},\nu)$ through the absorption 
coefficient $a(\nu)$ and the diffusion coefficient $\sigma(\nu)$ . The source terms 
are described by introducing the thermal emission $\eta(\nu)$, which depends at each 
point on the local temperature according to the Kirchhoff's law: 
$\eta(\nu) = a(\nu) B(\nu,T)$, and a diffusion (scattering) term $\sigma(\nu) J(\nu)$.
The symbol $J(\nu)$ denotes
the monochromatic mean intensity of the radiation field, {\it{i.e.}}
the integral of $I({\bf{n}},\nu)$ over all the directions ${\bf{n}}$ times $1/4 \pi$. 
In the processes
of scattering the amount of monochromatic radiative energy gained compensates exactly
the energy lost, when all the directions are taken into account.

Under the assumption of radiative equilibrium (RE), the constraint of energy balance reads

\begin{equation}\label{eq:1}
\int_{0}^{\infty}\ a(\nu) B(\nu,T)\ d\nu\ =\ \int_{0}^{\infty}\ a(\nu) J(\nu)\ d\nu\ ,
\end{equation}

which expresses the equality between thermal emission and the absorption processes. 
The rhs of eq. (\ref{eq:1}) is straightforwardly computed from the solution of the RT
equations. If this term is known, eq. (\ref{eq:1}) becomes an implicit transcendental 
equation that yields at each point of the atmosphere the local value of $T(r)$ in a 
trivial way. It must be stressed that at this stage of the procedure the absorption 
coefficient $a(\nu)$  is assumed to be known. The only difficulty is that $J(\nu)$
comes from the solution of the RT equations, which depends on the temperature through
the Planck function $B(\nu,T)$ that enters into the relative source terms.

This dependence entails introducing a new iterative procedure, which can be sketched as
follows. We start from a trial temperature distribution $T^{\ i} \equiv T^{\ i}(r)$,
which allows us to solve explicitly the RT equations, namely to get the specific 
intensities  $I({\bf{n}},\nu)$ at each point. We can then compute the corresponding 
mean intensity $J^{\ i}(\nu)$, from which we obtain the explicit value of the rhs of 
eq. 
(\ref{eq:1}), corresponding to the current value $T^{\ i}$ of the temperature. By 
solving eq. (\ref{eq:1}) as a transcendental equation we get the new temperature 
distribution $T^{\ i+1}$. Unfortunately the rate of convergence of this simple scheme, 
usually called the $\Lambda$-iteration, is exceedingly slow; in the numerical practice
often it does not converge at all.

Although the $\Lambda$-iteration does not converge, the relevant working plan is so 
simple that it deserves to be kept. A slight alteration of the above iterative 
procedure can dramatically improve the rate of convergence. Moreover - which is even 
more remarkable - that neither implies a heavy computational burden nor 
introduces the risk of numerical instabilities. The change has to be introduced 
once $J^{\ i}(\nu)$, the frequency integrated mean intensity $J^{\ i}$ and the 
total energy absorbed $J^{\ i}_{a}$ have been computed from the RT  equations.
The improvement consists
in the re-computation  of the total energy absorbed, whose value at each point we will
denote now by $J_{a}^{\ i\ c}$. These values are evaluated straightforwardly from the 
system of moments of the RT equation, and computed
in such a way that they meet certain properties of the converged solution. These 
properties, relevant to radiative transfer, are a consequence of the constancy of the total
flux $H$, {\it{\ i.e.}} the integral over all the frequencies of the monochromatic flux 
$H(\nu)$.

Let us consider in detail this new intermediate step. The RT equation for a one-dimensional planar medium is

\begin{equation}\label{eq:2}
\mu\ {\frac{d}{dr}} I(\mu,\nu)\ =\ \chi(\nu)\ \left[ I(\mu,\nu)\ -\ S(\nu)\right]\ ,
\end{equation}

where $\mu$ is the cosine of the angle between the direction of propagation 
${\bf{n}}$ and the $r$-axis, and $\chi(\nu) \equiv a(\nu) + \sigma(\nu)$ the total 
opacity. The source function  is expressed as

\begin{equation}\label{eq:3}
S(\nu)\ =\ \varepsilon(\nu) B(\nu,T)\ +\ \left[ 1\ -\ \varepsilon(\nu) \right]\ 
J(\nu)\ ,
\end{equation}

where  $\varepsilon(\nu)$ is defined as

\begin{equation}\label{eq:4}
\varepsilon(\nu)\ \equiv\ {\frac{a(\nu)}{a(\nu) + \sigma(\nu)}}\ =\ 
{\frac{a(\nu)}{\chi(\nu)}}\ .
\end{equation}

The equations for the three first $\mu$-moments of the specific intensity:
$J(\nu),\ H(\nu)$ and $K(\nu)$, are obtained by integrating eq. (\ref{eq:2}) over 
$d\mu$ and $\mu d\mu$ respectively:

\begin{equation}\label{eq:5}
{\frac{d H(\nu)}{d r}}\ =\ -\ a(\nu)\ \left[ J(\nu)\ -\ B(\nu,T) \right]
\end{equation}

and

\begin{equation}\label{eq:6}
{\frac{d K(\nu)}{d r}}\ =\ -\ \chi(\nu)\ H(\nu)
\end{equation}

By integration over all the frequencies, we get the bolometric equations

\[
\ \ \ \ \ \ \ \ \ \ \ {\frac{d H}{d r}}\ =\ B_{a}\ -\ J_{a}\ \ \ \ \ \ \ \ (7)
\ \ \ \ \ \ \ \ \ \ {\rm{and}}\ \ \ \ \ \ \ \ \ \ \ \ \ 
{\frac{d K}{d r}}\ =\  -\ H_{\chi}\ ,\ \ \ \ \ \ \ \ \ {\rm{(8)}}
\]

where $J,\ H$ and $K$ are the frequency-integrated zero- , first- and second-order
$\mu$-moments of the specific
intensity. The quantities $B_{a}$, $J_{a}$ and $H_{\chi}$ are
the integrals over all the frequencies of $B(\nu,T)$ and  $J(\nu)$ weighted
with the absorption coefficient $a(\nu)$, and that of  $H(\nu)$ weighted with the 
opacity $\chi(\nu)$, respectively.

We recognize in eq. (7) the statement of the energy balance: the difference
between the total emission and the total absorption is compensated by the variation
of the total radiative flux. Under the condition of RE, this difference must be 
zero, therefore the variation of H will be null. The constant total radiative flux 
$H^{\ast}$ that corresponds to the situation of RE is directly related to the 
luminosity, {\it{viz}} the effective temperature $T_{eff}$  of the star:

\setcounter{equation}{8}
\begin{equation}\label{eq:9}
H^{\ast}\ =\ {\frac{\sigma_{R}}{4 \pi}}\ T^{4}_{eff}\ ,
\end{equation}

where $\sigma_{R}\ =\ 5.6696 \cdot 10^{-5}\ erg \cdot cm^{-2} \cdot s^{-1} \cdot
K^{-1}$  is  the Stefan-Boltzmann constant.

Once introduced these preliminary definitions, we can get back to the 
iterative correction of the temperature. In the first part of each step of 
iterations, given the current value of the temperature $T^{\ i}$ and consequently that of
$B(\nu,T^{\ i})$, we can compute not only $J^{\ i}(\nu)$, but also 
$H^{\ i}(\nu)$ and $K^{\ i}(\nu)$; hence, trivially, the frequency-integrated 
quantities: 
$J^{\ i}$,
$J_{a}^{\ i}$, $H^{\ i}$, $H_{\chi}^{\ i}$ and $K^{\ i}$. 
The extra computational work for these quadratures
adds very little to that required for the quadrature of $J(\nu)$. Then, by making 
use of the foregoing quantities, we compute at each point the following ratios:

\[
\ \ \ f\ \equiv\ {\frac{K^{\ i}}{J^{\ i}}}\ \ \ \ \ \ \ (10)\ ,\ \ \ \ \ 
\beta\ \equiv\ {\frac{H^{\ i}_{\chi}}{H^{\ i}}}\ \ \ \ \ \ \ (11)\ \ \ \ \ 
{\rm{and}}\ \ \ \ \ a_{J}\ \equiv\ {\frac{J^{\ i}_{a}}{J^{\ i}}}\ \ \ \ \ \ \ 
(12)\ .
\]
 
Because these ratios are the quotient between homologous quantities (in the practice
pairs of quantities with the same physical behavior), they shall result almost 
independent of the input $B(\nu,T^{\ i})$. We can consider them as 
{\it{quasi-invariant}},
and employ them as iteration factors.

Afterward we have obtained - in such an easy way - the iteration factors, we 
will
proceed recomputing the improved value of the energy absorbed: $J^{\ i\ c}_{a}$. 
Thus, by recalling the definition of the factor $\beta$ given by eq. (11), we can 
rewrite eq. (8) in the form

\setcounter{equation}{12}
\begin{equation}\label{eq:13}
{\frac{d K}{d r}}\ =\ -\ \beta H^{\ast}\ ,
\end{equation}

where we have introduced the value of  $H^{\ast}$, a data that characterizes the 
problem under consideration, as shown by eq. (\ref{eq:9}). The quasi-invariant 
iteration factor $\beta$ is almost independent of  the input $B(\nu,T^{\ i})$. 
That allows
us to obtain an improved value $K^{\ i\ c}$ for the moment $K$, which takes into 
account the correct value of the radiative flux. The first iteration factor $f$, 
defined by eq. (10), is the Variable Eddington Factor. It shall yield the 
corresponding improved value of  $J^{\ i\ c}$.

The third quasi-invariant factor  $a_{J}$, defined by eq. (12), shall furnish us 
the up-dated value of the total energy absorbed, {\it{i.e.}} 
$J_{a}^{\ i\ c}\ =\ a_{J}\ J^{\ i \ c}$.  This is the rhs of the equation 
that expresses the constraint of 
RE. We will solve the transcendental equation
 
\begin{equation}\label{eq:14}
\int_{0}^{\infty}\ a(\nu) B(\nu,T^{\ i+1})\ d\nu\ =\ J_{a}^{\ i\ c}\ , 
\end{equation}             

by using now the improved value  $J_{a}^{\ i\ c}$,  in order to get the up-dated 
temperature  $T^{\ i+1}$ at each point.

On  one hand the iteration factors are almost independent  of the input 
$B(\nu,T^{\ i})$; on the other they are very easily computed from the RT solution. Thus 
the computation of the improved value  $J_{a}^{\ i\ c}$ of the total energy absorbed is
as simple in the practice as the previous computation of $J_{a}^{\ i}$ directly from the 
RT solution.
But now $J_{a}^{\ i\ c}$ includes the information on the constancy of the total 
radiative flux  $H^{\ast}$, namely the condition of RE: 
{\it{by employing the iteration 
factors we solve the set of the RT equations consistently with the constraint of 
energy conservation}}. That makes the iterative procedure extraordinarily fast: the 
rate of convergence improves dramatically at the price of a little extra 
computational cost.

\section{A Pending Difficulty with the Temperature Correction}
\label{sec:3}

It is well known , however, that within each one of the integrals in eq. 
(\ref{eq:1}) the contributions due to the spectral ranges corresponding to the most
opaque radiative transitions can overwhelm by  many orders of magnitude those 
due to the remaining frequencies. On the other hand, far from the superficial layers
the monochromatic mean intensity $J(\nu)$
will be equal up to many significant digits to the source function $S(\nu)$, and 
consequently to the Planck function  $B(\nu,T)$, at those frequencies for which the 
absorption coefficient $a(\nu)$  - hence the  opacity $\chi(\nu)$ - is very large. 
Then the net contribution of the latter to the energy balance must be null, albeit 
separately their contribution to the two integrals are far the most important 
numerically. Thus those spectral ranges, whose contribution to the overall balance 
is null, will dominate numerically both side of the relevant equation, and 
consequently the errors on the determination of  $J(\nu)$ and  $S(\nu)$ at these 
frequencies shall falsify the balance.

A way to circumvent this severe numerical problem is to solve the RT equations in 
the new variable $I({\bf{n}},\nu) - S(\nu)$, instead of the specific intensity 
$I({\bf{n}},\nu)$ customarily employed. Then it will be possible to compute the 
differences $J(\nu) - B(\nu,T)$ 
directly from the solution of the RT equations. The  spurious 
contributions above mentioned disappear, because the differences automatically 
vanish when it is the case, independently of the absolute vales of  
$I({\bf{n}},\nu)$ and  $B(\nu,T)$. The introduction of the new variable 
$I({\bf{n}},\nu) - S(\nu)$ comes in a natural way from the use of our Implicit 
Integral Method (see~\cite{Simonneau93} and~\cite{Crivellari94}). 
Its application to the problem under study here have been preliminary 
discussed in~\cite{Cardona02}, both from the physical and the mathematical standpoint. 

By combining the above way of solving the RT 
equations with the method of the iteration factors, 
it is possible to correct the temperature in spite of the difficulties mentioned previously.
In the first part of each step 
of iterations, given the input  $B(\nu,T^{\ i})$, we will compute directly the differences 
$J(\nu) - B(\nu,T^{\ i})$ for each frequency. Hence, by adding to the latter the data 
$B(\nu,T^{\ i})$, we will obtain the monochromatic mean intensities $J(\nu)$, as well as
all the other moments, both monochromatic and frequency-integrated, that we need in 
order to compute the iteration factors defined by eqs. (10) through (12). It must be
stressed that we get directly the differences $J(\nu) - B(\nu,T^{\ i})$, and 
consequently their product by the absorption coefficient $a(\nu)$ integrated over 
all the frequencies, {\it{i.e.}}

\begin{equation}\label{eq:15}
\left( J\ -\ B\right)_{a}^{\ i}\ =\ \int_{0}^{\infty}\ a(\nu)\ \left[ J(\nu)\ -\ 
B(\nu,T^{\ i}) \right]\ d\nu\ ,
\end{equation}

is evaluated without the need of computing separately  its two components 
$J_{a}^{\ i}$ 
and $B_{a}^{\ i}$.  If $(J - B)_{a}^{\ i}$ is null at each depth point, the condition 
of RE is fulfilled everywhere, and therefore the process of correcting the temperature 
is over. If not, we must proceed with the second part of the iteration step to 
determine a new temperature distribution  $T^{\ i+1}$, via an equation like eq. (15).

Once we have obtained the values of $(J - B)_{a}^{\ i}$ on the one side, and the 
monochromatic and frequency-integrated moments (in particular $J_{a}^{\ i}$) as well as 
the iteration factors on the other, we go forward to compute the improved values  
$J^{\ i\ c}$ and $J_{a}^{\ i\ c}$. As already said before, on one hand the dependence 
of the latter on the input  $B(\nu,T^{\ i})$ is weaker than that of  $J_{a}^{\ i}$, on 
the other they take into account the information on the constancy of the total 
radiative flux  in a stronger way. The effect brought about by these advantages is 
measured by the factor

\begin{equation}\label{eq:16}
\gamma\ \equiv\ {\frac{J^{\ i\ c}}{J^{\ i}}}\ =\ {\frac{J_{a}^{\ i\ c}}{J_{a}^{\ i}}}\ ,
\end{equation}

whose value converges very quickly to unit.

Let us get back now to eq. (\ref{eq:15}), which is the protagonist of the second 
part of each step of iterations. We can add to each of his members the quantity 
$\gamma B_{a}(T^{\ i})$, where $B_{a}(T^{\ i})$ is the input for the current step of 
iterations, in order to get

\[
\int_{0}^{\infty}\ a(\nu)\ \left[ B(\nu,T^{\ i+1})\ -\ \gamma B(\nu,T^{\ i}) \right]\
d\nu\ =\ J_{a}^{\ i\ c}\ -\ \gamma B_{a}(T^{\ i})\ =\]
\begin{equation}\label{eq:17}
=\ \gamma\ \left[ J_{a}^{\ i}\ -\ B_{a}(T^{\ i}) \right]\ =\ \gamma \left( J\ -\ B
\right) _{a}^{\ i}\ .
\end{equation}

The term  $(J - B)_{a}^{\ i}$, defined by eq. (\ref{eq:15}), is actually what we 
compute directly as a whole, starting from the solution $I({\bf{n}},\nu) - S(\nu)$
of the RT equations. In such a way there vanish automatically the differences 
between absorption and emission for those frequencies, whose net contribution to 
the energy balance must be null, in spite of the fact that the corresponding values
of $a(\nu) J(\nu)$  and $a(\nu) B(\nu,T)$ are overwhelming (sometimes by  many 
orders of magnitude) in the corresponding integrals.

That was exactly our aim. The solution of the new transcendental equation (eq. 
[\ref{eq:17}]), that we 
have to solve now, presents the same degree of difficulty as the previous equation 
(\ref{eq:14}). We can conclude that to iterate by making use of the iteration 
factors not only speeds up dramatically the rate of convergence of the procedure, 
but also allows us to take into in account in the numerical algorithm only
the intrinsic effects of the physical process that are actually necessary.
       
The term  $(J - B)_{a}^{\ i}$ represents a local correction:  at those points where it
is different from zero the condition of RE is not satisfied, and consequently it 
will be necessary to correct the temperature there. However it may occur that 
$(J - B)_{a}^{\ i}$ be null at a certain point, and nevertheless the correction be 
necessary, because the condition of RE is not fulfilled elsewhere. This non-local 
effect is taken into account (iteratively) through the factor $\gamma$.

\section{Conclusions}
\label{sec:4}

We have presented here a novel method for building stellar atmosphere models, more precisely,
for the calculations required by the determination of the temperature at each point.
Because of radiative transfer one must take into account the coupling among the physical 
conditions at each and every point of the system. In particular, the temperature at each
point is linked with those at all the others.
We ideally remove the coupling, and proceed
sequentially to get the physical conditions at each point under the assumption that those at
all the other points be known. In order to achieve the consistency - {\it{i.e.}} the 
coupling - we developed an iterative procedure. Such a treatment constitutes a numerical 
simulation of the physical processes that are at the origin of the corresponding equations.
We tackle and solve those problems that arise when the fluxes to be determined
are the difference between quantities (densities) that are very close, so that the 
difference between their values is less than the numerical accuracy of the computations.

As a proof of the quality of the method proposed, we claim that the integrated radiative
flux $H^{\ i}$, computed at each step of iterations, quickly converges (in a few iterations)
to the actual value $H^{\ast}$. The inaccuracies on the converged
numerical values of $H^{\ i}$ are
due to the quadrature of $H^{\ i}(\nu)$ over frequencies, inaccuracies that we
estimate to be more close to $0.01\%$ than to $0.1\%$. 
As the problem with the numerical quadratures over frequencies is the same as above, we
impose that the equality between the two terms of the transcendental equation 
(\ref{eq:14}) be numerically accurate to one order of magnitude less, namely to
$0.001\%$, in order not to increase the unavoidable indetermination already
brought about by the numerical treatment of the data.


\begin{thebibliography}{99.}
\bibitem{Simonneau88}Simonneau, E. and Crivellari, L.: 1988, Astrophys. J., 330, 415.
\bibitem{Crivellari91}Crivellari, L. and Simonneau, E.: 1991, Astrophys. J., 367, 612.
\bibitem{Simonneau93}Simonneau, E. and Crivellari, L.: 1993, Astrophys. J., 409, 830.
\bibitem{Crivellari94}Crivellari, L. and Simonneau, E.: 1994, Astrophys. J., 423, 331.
\bibitem{Cardona02}Cardona, O., Crivellari, L. And Simonneau, E.: 2002, New Quests in Stellar 
Astrophysics: The Link between Stars and Cosmology, (Dordrecht: Kluwer Academic 
Publishers), p. 29.
\end{thebibliography}
\end{document}